\documentclass{ws-ijmpa}

\begin{document}

\markboth{Pasquale Blasi}{On the Origin of Very High Energy Cosmic Rays}

%
%

\title{\bf ON THE ORIGIN OF VERY HIGH ENERGY COSMIC RAYS}

\author{PASQUALE BLASI}

\address{INAF/Osservatorio Astrofisico di Arcetri\\
Largo E. Fermi, 5 - I50125 Firenze (ITALY)}

\maketitle

\begin{history}
\received{Day Month Year}
\revised{Day Month Year}
\end{history}

\begin{abstract}
We discuss the most recent developments in our understanding of the 
acceleration and propagation of cosmic rays up to the highest energies. 
In particular we specialize our discussion to three issues: 1)
developments in the theory of particle acceleration at shock waves;
2) the transition from galactic to extragalactic cosmic rays; 3) 
implications of up-to-date observations for the origin of ultra high
energy cosmic rays (UHECRs).
\end{abstract}

\section{Introduction}

The spectrum of cosmic rays observed at the Earth between $10^{11}$ eV 
and $10^{21}$ eV is illustrated in Fig. 1: at
energies below $100$ GeV the observed spectrum is heavily affected 
by local phenomena such as the interaction with the solar wind and the
magnetic field of the Earth.
In Fig. 1, the flux has been multiplied by $E^3$
in order to emphasize the possible departures from a power law $E^{-3}$.
There are basically three features that may be identified in such 
spectrum: 1) a knee at energy $E_K\approx 3\times 10^{15}$ eV, 
consisting of a steepening of the all-particle spectrum
from $E^{-2.7}$ to $\sim E^{-3}$; 2) a second knee at energy 
$E_{2K}\approx 10^{18}$ eV; 3) a dip at $E_D\approx 5\times 10^{18}$ eV. While 
all of these features result from observations, a fourth feature has
been predicted to exist on pure theoretical grounds, though it has
not been observed yet, mainly because of the insufficient statistics 
of collected events: 
at energies $\ge 10^{20}$ eV, the photopion production interactions of 
protons with the cosmic microwave background should produce 
a suppression in the flux, the so-called {\it Greisen-Zatsepin-Kuzmin 
(GZK) feature}\cite{gzk}. The detection of such feature would
represent the best proof that these cosmic rays are of extragalactic origin. 

While the detection of the GZK feature has been and in fact is one of
the most important goals for cosmic ray physics, for the implications
that it has on both physics and astrophysics, the problem of the
origin of cosmic rays as a whole includes many more issues and open
problems. It appears clear from recent findings that the collection of 
high statistics of events and credible measurements of the chemical 
composition are crucial ingredients for the understanding of the 
phenomenon at all energies. At energies above $\sim 10^{17}$ eV, the
measurements of anisotropy also become a very precious tool. 

In this review we limit our discussion to three issues that 
appear particularly important at the present time: 
1) the acceleration of cosmic rays 
through diffusive motion across shock fronts (Sec. \ref{sec:shocks}); 
2) the transition from cosmic rays accelerated within our Galaxy to 
those that reach us from extragalactic sources (Sec. \ref{sec:transition}); 
3) the spectrum and small scale anisotropies of UHECRs (Sec. \ref{sec:gzk}).
We summarize in Sec. \ref{sec:summary}.

\begin{figure}[thb]
 \begin{center}
  \mbox{\epsfig{file=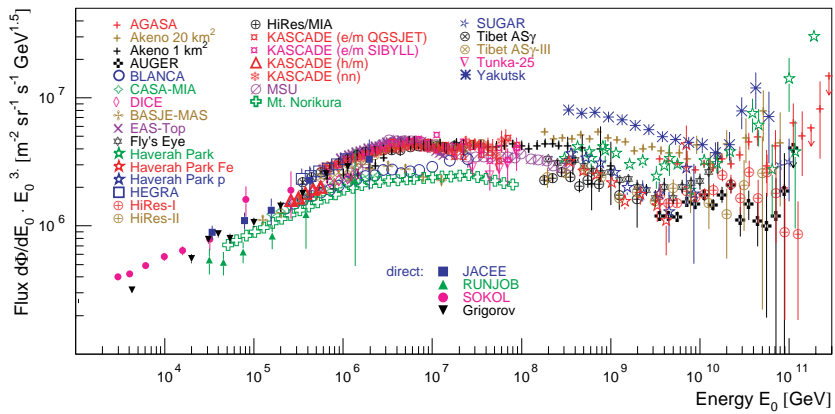,width=11.cm}}
  \caption{All-particle spectrum of cosmic rays as observed at the Earth
  (collection of data from several experiments as in Ref. $^{20}$. The
  references to the experiments listed in the figure can also be found
  in Ref. $^{20}$.)}
 \end{center}
\label{fig:spectrum} 
\end{figure}

\section{Diffusive acceleration of cosmic rays at collisionless shocks}
\label{sec:shocks}

\subsection{Non-relativistic shocks: a paradigm confronting observations}
\label{sec:shocks1}

The paradigm that is generally invoked to explain the origin of 
the bulk of galactic cosmic rays is based on the acceleration of 
particles at collisionless non-relativistic shock waves that 
develop in the supersonic motion of the ejecta of supernova
explosions \cite{ginz} (see \cite{hillas} for a recent excellent
review and \cite{bible} for a broader view of the subject). 
Particle acceleration at similar shocks in other sources 
(e.g. active galaxies, radio lobes) is often invoked to explain the
origin of higher energy cosmic rays.

Building on the original idea of Fermi \cite{fermi}, in the 70's a 
deep investigation of stochastic acceleration in the presence of shock
waves was started \cite{krymsky,bo,bell78} (see
\cite{drury83,bl_ei,je91} for reviews). 
The scattering of charged particles back and forth
through the shock front was shown to lead to energization of the
particles. The advection of the accelerating particles with the fluid
towards downstream infinity is responsible for the formation of a
spectrum which has the form of a power law in momentum. In the case of
strong shock waves (Mach number $M\gg 1$) the spectrum tends
asymptotically to have a spectrum $\propto E^{-\gamma}$ with slope 
$\gamma \sim 2$ \cite{krymsky,bo}
\footnote{Here and in the following we will use the expression {\it slope
  $\gamma$ of the spectrum}  when we refer to the distribution function
  in energy space, so that the total number of particles is given 
  by $\int dE f(E)$. We will use the symbol $\gamma_p$ or $s$ to refer to
  the slope of the distribution function of accelerated particles 
  in momentum space, normalized in such a way that the number of 
  particles is $\int dp 4\pi p^2 f(p)$. Clearly $\gamma_p=2+\gamma$
  for ultra relativistic particles.}.

For cosmic rays that are originated in the Galaxy and reach the Earth
through a diffusive wandering motion, the spectra get modified 
by diffusion, so that an injection spectrum $Q(E)\propto E^{-\gamma}$ is 
steepened to an equilibrium spectrum $n(E)\approx Q(E)\tau_{dif}(E)\propto 
E^{-(\gamma+\alpha)}$, where $\tau_{dif}(E)\propto 1/D(E)$ is the diffusion time 
for particles with energy $E$ and $D(E)\propto E^{\alpha}$ is the 
diffusion coefficient. Energy losses and turbulent reacceleration 
(namely second order Fermi acceleration) can slightly but not 
substantially affect this result at energies larger than $\sim 100$
GeV \cite{turbul}. Fragmentation of heavy nuclei changes the slope 
of the observed spectra for those chemical elements for which these 
processes are relevant (most notably iron). 
The diffusion coefficient in the interstellar medium (ISM) can be 
inferred from the abundances of light elements (primarily Lithium, Boron 
and Berillium, scarse in the primordial soup and not appreciably
synthesized in stars), produced almost entirely 
through spallation of heavier elements (see ref. \cite{spalla} and
references therein). From the measurement 
of the abundance of these light elements (as well as from other 
observational constraints, widely discussed in \cite{schlick} and
references therein) it is 
possible to infer an estimate of the typical time of residence of 
cosmic rays with energy $\sim 1$ GeV in the Galaxy. This time turns
out to be of several million years, the best proof that cosmic rays
diffuse in  the Galaxy (the crossing time for straight line
propagation would be only $\leq 3\times 10^4$ years). 

For our purposes we can adopt as an order of magnitude for the
diffusion coefficient in the Galaxy $D_{gal}(E)\approx 3\times 
10^{29}(E/GeV)^\alpha~\rm cm^2 s^{-1}$, with $\alpha \approx 0.6$ 
($\alpha=1/3$ would correspond to Kolmogorov spectrum of fluctuations
and would imply a somewhat smaller normalization. $\alpha=0.5$ corresponds
to Kraichnan diffusion). The observed spectrum at energies below the
knee forces $\gamma+\alpha = 2.7$ while leaving the two slopes
separately unconstrained. If indeed $\alpha=0.6$, the injection 
spectrum would be forced to be $\propto E^{-2.1}$ \footnote{It is
not clear if this slope of the diffusion coefficient is compatible
with the observed anisotropy above the knee (see \cite{hillas} for
details and additional references).}

The maximum energy of the accelerated particles is the main concern
for the scenario just outlined: if the supernova explosion occurs
in the ISM, the diffusion coefficient that determines the motion of 
the particles upstream of the supernova shock is of the order of 
$D_{gal}(E)$ (see above). 
The acceleration time is therefore $\tau_{acc}(E)\approx
D_{gal}(E)/u_{sh}^2$, where $u_{sh}$ is the shock velocity 
($\sim \rm a~few~1000~km/s$). For a supernova remnant of
typical age $\tau_{SNR}\sim 1000$ years, the maximum energy of protons 
is easily estimated by requiring that the acceleration time remains
smaller than $\tau_{SNR}$, and it is found to be of 
the order of {\it fractions of GeV}. The maximum energy of electrons
is determined by the balance between acceleration and energy losses. 

This estimate strongly suggests that acceleration of protons is
possible only if the diffusion coefficient close to the shock is much
smaller than that in the interstellar medium. One realistic way to 
achieve this condition is to allow for the self-generation of magnetic
turbulence by the accelerated particles \cite{bell78}: 
if charged particles move in a magnetized medium with a bulk velocity 
that exceeds the Alfv\`en speed, the {\it streaming instability}
is excited, thereby limiting the speed of the bulk of cosmic rays 
to roughly the Alfv\`en speed. 
In this case the diffusion coefficient upstream of the shock is
determined by the density of accelerated particles. In the context of
quasi-linear theory the form of the diffusion coefficient can be
written as $D(E)=D_B(E)/\cal{I}$, where $D_B(E)=(1/3)r_L(E)c$ is 
the Bohm diffusion coefficient ($r_L$ here is the Larmor radius of 
particles with energy $E$). $\cal I$ is the energy density of Alfven 
waves resonant with particles with energy $E$ relative to the
background magnetic field $B$:
\begin{equation}
\frac{\delta B^2}{8\pi} = \frac{B^2}{8\pi} \int \frac{dk}{k} {\cal I}(k).
\label{eq:def}
\end{equation}

In terms of quasi-linear theory, the value of $\cal I$ is bound to
be less than unity. If ${\cal I}=1$ (strong turbulence), then diffusion 
occurs in the the Bohm regime \cite{bell78,lc83}. 
In \cite{lc83,lc83_2} the authors
demonstrate that the maximum energy of protons within this scenario 
is as large as $E_{max}\approx 10^{13}-10^{14}$ eV, and $Z$ times 
larger for heavier nuclei with charge $Z$. Despite 
this improvement, the $E_{max}$ obtained in \cite{lc83} is still $\sim
30$ times smaller than the energy at which the knee is measured
(see Fig. 1).

This relatively low value of $E_{max}$ is the direct consequence of 
assuming that the saturation of the self-generated
turbulence occurs when the linear theory breaks down, namely when
${\cal I}\sim 1$. This might be the case if the fraction of energy 
channelled into cosmic rays is small, $P_{CR}\ll \rho u^2$.
On the other hand, the maximum theoretical saturation level that can
be predicted, even in the context of quasi-linear theory, is much
higher, as was found in \cite{vmc82}, by simply using the basic equation 
that describes the amplification and convection of waves:
\begin{equation}
u\frac{d {\cal F}(k)}{d x} = v_A \frac{d {\cal P}_{CR}}{d x},
\label{eq:growtheq}
\end{equation}
where $u$ is the shock speed, $v_A$ is the Alfv\`en speed
(calculated using the background unperturbed magnetic field)
and ${\cal F} = \frac{B^2}{8\pi} \cal I$ is the energy density 
in the waves with wanumber $k$
\footnote{The correct equation should in fact be 
$$\frac{\partial {\cal F}}{\partial t} + 
(u+v_A)\frac{\partial {\cal F}}{\partial x} 
= v_A \frac{d {\cal P}_{CR}}{d x}.$$ Here we are assuming that a stationary
situation is reached and that the fluid is strongly super-alfvenic,
$u\gg v_A$.}. The pressure ${\cal P}_{CR}$ in Eq. \ref{eq:growtheq} is the
one that refers to particles with momenta $p$ that can resonate with
waves with wavenumber $k$, namely those for which $k^{-1}=p|\mu|/ZeB$,
where the wavenumber $k$ is assumed to be parallel to the direction of
the background magnetic field and $\mu$ is the pitch angle of the
particle. If one assumes a sharp resonance at $p=p_{res}=ZeB/ck$, then
$${\cal P}_{CR} = \frac{4\pi}{3}p_{res}^4 v(p_{res}) f(p_{res}).$$
If one is interested in the total magnetic field amplification $\delta
B^2$, as defined in Eq. \ref{eq:def}, then the previous equation
becomes
\begin{equation}
u\frac{d}{d x}\frac{\delta B^2}{8\pi} = v_A \frac{d P_{CR}}{d x},
\label{eq:growtheq1}
\end{equation}
where now $P_{CR}$ is the total cosmic rays pressure at the shock.

Integration of this equation implies that
\begin{equation}
\frac{\delta B^2}{B^2} = 2 M_A \frac{P_{CR}}{\rho u^2},
\label{eq:saturation}
\end{equation}
where $M_A$ is the Alfv\`en Mach number. If $P_{CR}\sim \rho u^2$,
the amplification of the magnetic field with respect to the background
undisturbed field can be as high as $M_A^{1/2}$. Several processes can 
reduce the effective magnetic field to values much lower than those 
found through Eq. \ref{eq:saturation}\cite{ptu}. On the other hand,
\ref{eq:saturation} immediately shows that the linear theory that it 
is based upon is easily broken in the description of efficient
particle acceleration at shocks.

Some authors have argued that shock acceleration in the denser
environment of some pre-supernova environments could help reaching
energies higher than or comparable with those predicted by
\cite{vb,bier93}. This scenario may take place in some sources and 
may work together with the process of self-amplification of the 
magnetic field explained above. 

The concern about the highest energy of particles accelerated at 
SNRs has become even more serious with the recent findings of the 
KASCADE experiment that could measure the spectra of different
chemical elements separately. 
A collection of the data for protons and helium nuclei from several 
experiments is shown in Fig. 2 (from Ref. \cite{hora}). For 
heavier elements the uncertainties in the data become large 
and prevent to reach useful conclusions at the present time. 
\begin{figure}[thb]
 \begin{center}
  \mbox{\epsfig{file=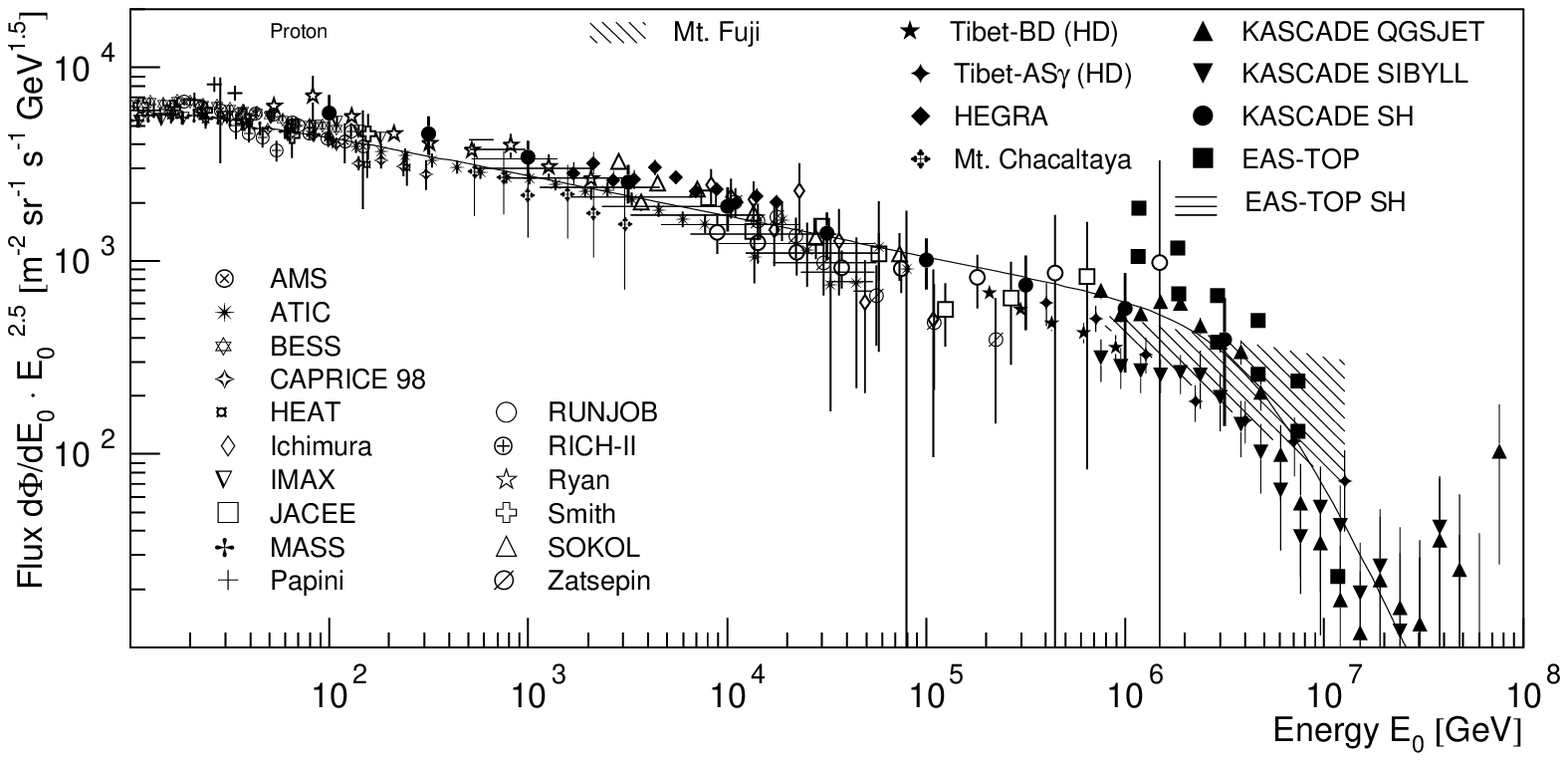,width=6.cm}}
  \mbox{\epsfig{file=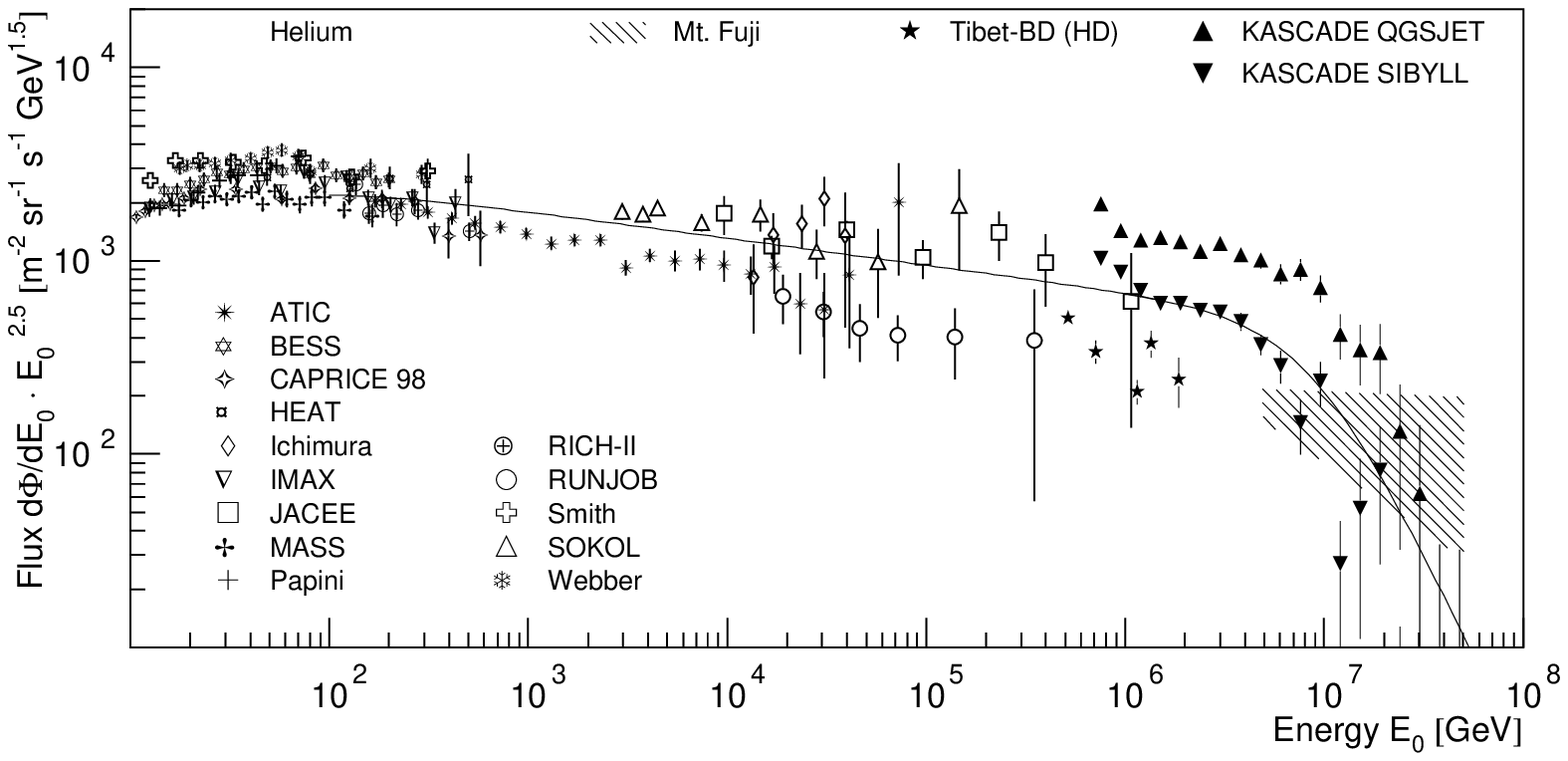,width=6.cm}}
  \caption{Spectrum of protons (left panel) and helium nuclei (right
    panel) as reported in Ref. $^{20}$.}
 \end{center}
\label{fig:compo}
\end{figure}
For our purposes the continuous lines \cite{poli} can be used just to guide the
eye. The KASCADE data \cite{kascade} suggest the intriguing
possibility that the observed knee in the all-particle spectrum may 
result from the superposition of different power laws, each describing
the spectrum of different chemical components, and each with a
rigidity dependent knee. If this interpretation is confirmed, then the
{\it observed} knee in the all-particle spectrum may not have direct
physical meaning and the knees in the single components would require
an explanation, either in terms of acceleration in the sources or in
terms of propagation in the Galaxy (see Ref. \cite{luke} for an
attempt at explaining these knees in terms of acceleration in SNRs). 

Two pieces of information can be inferred from Fig. 2: 
1) protons are measured
up to energies $(2-4)\times 10^{16}$ eV, that exceeds the maximum energy
predicted by standard diffusive shock acceleration by at least two 
orders of magnitude. Helium nuclei are measured up to energies roughly 
twice as large as those of protons, as would be expected on the basis of a
rigidity dependent process; 2) Protons and Helium each seem to have their 
own rigidity dependent {\it knees}. The spectrum above these knees is 
characterized by a steepening rather than a cutoff (the approximate 
slopes are $\sim 2.7$ below the knees and $\sim 3.8$ above). Following 
the same rigidity criterion, iron nuclei would have their knee at 
$E_{Fe}\sim 4\times 10^{16}$ eV and maximum energies at $\ge 10^{17}$ eV.
Unfortunately no solid conclusion on heavy nuclei can be reached
from the data at the present time. 

We will discuss the possible implications of point 2) in Sec. 
\ref{sec:transition}. Here we wish to discuss the very basic issue
of the maximum energy of particles accelerated at shock waves.
Although the main conclusions will apply to shocks in SNRs, 
similar considerations might be reached, on the case by case 
basis, for sites of acceleration of UHECRs where shocks are
invoked.

We have already pointed out that the lineary theory of particle
acceleration at shock waves fails to explain maximum energies as high
as that of the knee by a factor $\sim 30$ and it fails to explain
proton energies as high as those measured by KASCADE and shown in 
Fig. 2 by a factor $\sim 100$. In the following we
shall describe how recent developments of non-linear theories of 
particle acceleration may change this conclusion dramatically.

One of the main achievements of recent investigations on particle
acceleration at collisionless shock waves consists of having 
introduced the reaction of the accelerated particles onto the
accelerating shock and the fluid itself. These modern theories 
no longer treat the accelerated particles as test particles. 
This phenomenon has received much attention in the context of the 
so-called two-fluid models \cite{dr_v80,dr_v81}, kinetic models 
\cite{malkov1,malkov2,blasi1,blasi2,vannoni,elena} 
and numerical approaches, both Monte Carlo and other numerical procedures 
\cite{je91,bell87,elli90,ebj95,ebj96,kj97,kj05,jones02}.
For a relatively recent accurate review see the work by
\cite{maldru2001}, which also contains a discussion of the role
of injection of protons and electrons. 

These independent approaches confirm that the cosmic ray reaction 
enhances the acceleration efficiency and at the same time flattens 
the spectra of particles at the highest energy end (close
to the maximum energy). Typical spectra and slopes of the 
accelerated particles are shown in Fig. 3
(from Ref. \cite{elena}) as obtained with an exact solution applicable to
arbitrary diffusion properties of the plasma: the upper panel shows 
the spectra obtained for maximum energy of $10^5$ GeV and Mach number of 
the shock as in the caption. The lower panel shows the local
slope as a function of momentum ($q(p)=-d\log f(p)/d\ln p$). 
The predicted spectra, which are no longer power laws, are typically 
softer than the linear prediction at low energy and harder (as hard 
as $f(p)\propto p^{-3.2}$) at the highest energies.

The concave spectra obtained in the context of non-linear particle
acceleration at shock fronts reflect the formation of a precursor in
the upstream fluid, namely a region with a size of the order of the
diffusion length of the highest energy particles, in which the density
and velocity profiles are space dependent. More specifically, the
fluid speed increases whereas density decreases while moving
away from the shock itself. This results in the appearance of a
compression factor that is a funtion of the distance from the 
shock: the total compression factor $R_{tot}$ between upstream infinity
and downstream can reach values much larger than the classical
strong shock limit, $R_{tot}\gg 4$, while for the compression factor at the
gaseous shock $R_{sub}\leq 4$. Particles with different momenta {\it
  feel} different compressions and this is reflected into the
appearance of non-power-law spectra, as shown in Fig. 3.
The shock modification that we just explained is somewhat reduced 
due to the transfer of energy from cosmic rays to the thermal gas,
as a consequence of Alfv\`en heating \cite{heating,vmc82} or Drury 
instability \cite{drury_ins}.

\begin{figure}[thb]
 \begin{center}
  \mbox{\epsfig{file=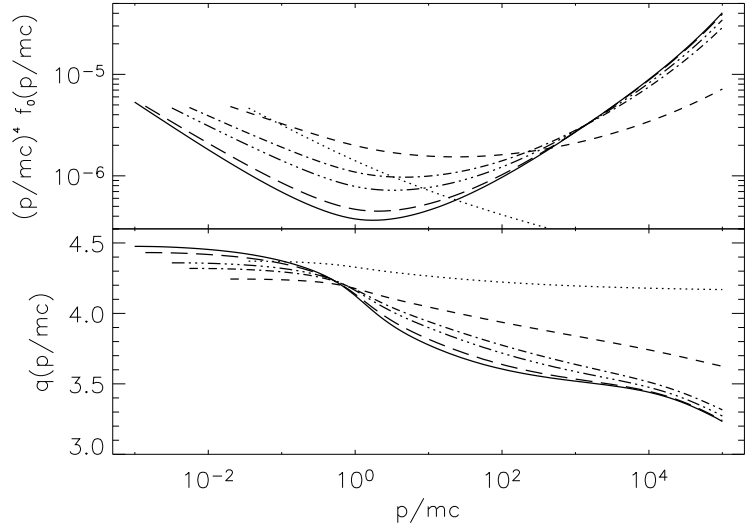,width=10.cm}}
  \caption{{\it Upper panel}: Spectra of accelerated particles at the 
location of the shock for $M_0=4$ (dotted line), 10 (short-dashed
line), 50 (dash-dotted line), 100 (dash-dot-dot-dotted line), 300 
(long-dashed line) and 500 (solid line).
{\it Lower panel}: momentum dependent slope for the same values of 
Mach numbers.}
\end{center}
\label{fig:nonlin}
\end{figure}

It is important to realize however that these spectra are 
{\it instantaneous spectra} and need to be convolved with the 
time evolution of the supernova parameters in order to provide 
the spectrum of cosmic rays originated from the supernova remnant 
of interest. The particles that are detected at the Earth as cosmic
rays are liberated from the downstream region of the SNR only after
the supernova shock dies out \footnote{In principle only particles with 
momenta close to $p_{max}$ can escape from the upstream region.
In the context of non-linear particle acceleration at shocks
however, these particles carry a non negligible fraction of the
total cosmic ray energy density}. Current calculations (not including
the magnetic field amplification induced by cosmic rays) predict 
curved spectra as a result of this temporal evolution (see for instance 
\cite{elli}). It is not clear if this curvature may be compatible 
with the steep power law spectra observed at the Earth, as it is not 
known how the magnetic field amplification at the shock location
during the supernova evolution may change this result.

A specific prediction of these nonlinear approaches is that it is
typical to have situations in which the pressure of cosmic rays becomes 
comparable with the total available kinetic energy of the fluid, 
$P_{CR}\sim \rho u^2$ and most of this pressure is actually
carried by the particles with the highest momenta, another important
difference with the case of acceleration in the linear regime. 

The fact that $P_{CR}$ and  $\rho u^2$ may be comparable changes 
considerably the development of the streaming instability with 
respect to the original calculations in \cite{bell78,lc83}. 
This has recently been investigated by \cite{bl00,bl01,bell2004}
who in fact claim that the background magnetic field close to
the shock may be amplified by a factor up to $\sim 10^3$. The most 
straightforward implication is that the corresponding maximum
energy of the accelerated particles is larger than found by 
\cite{lc83} by the same factor. It follows that the maximum 
energy predicted for protons would be $E_{max}^p \approx 10^{16}-
10^{17}$ eV, while being $Z$ times larger for nuclei with charge
$Z$. While there may be some effects that may saturate the 
instability to lower values of the turbulent magnetic field,
this remains certainly an exciting development that has the potential to
change our picture of particle acceleration in SNRs and in other 
scenarios as well, and in particular in the potential sources of ultra high 
energy cosmic rays. Another possible avenue to enhance the maximum
energy of accelerated particles at cosmic ray modified shocks has 
been recently proposed in \cite{diamond}.

The main result, discussed in detail in \cite{bell2004} is that the 
efficient particle acceleration is responsible for a new branch of 
purely growing non-alfvenic modes, in the quasi-linear regime. The 
growth of these waves is found to be faster than that of Alfvenic 
modes and may be responsible for saturation at a level which is 
appreciably higher than that found in Eq. \ref{eq:saturation}. In 
particular the saturation found by \cite{bell2004} with the help of 
numerical simulations is 
$\frac{\delta B^2}{8\pi} \approx \frac{u}{2c}P_{CR}$, namely
\begin{equation}
\frac{\delta B^2}{B^2} = M_A^2 \frac{u}{c} 
\frac{P_{CR}}{\rho u^2}.
\label{eq:sat_bell}
\end{equation}
By comparing this with Eq.\ref{eq:saturation}, we can see that the
really new result is the different saturation level of the
self-induced turbulence: for efficient particle acceleration, 
Eq. \ref{eq:saturation} provides $\delta B/B
\sim M_A^{1/2}$, while Eq. \ref{eq:sat_bell} gives the much higher 
$\delta B/B\sim M_A (u/c)$. One should notice that
Eq. \ref{eq:growtheq} cannot be applied in a straightforward way
to the unstable modes found in \cite{bell2004}, because the term 
on the right hand side of that equation only applies to the growth 
rate of Alfv\`en waves. 

It is worth stressing that the calculations of 
\cite{bl00,bl01,bell2004} are actually not currently carried out
in a self-consistent way: they all {\it assume} that $P_{CR}\sim 
\rho u^2$ but the calculations of the cosmic ray induced
instabilities are performed assuming power law spectra of accelerated
particles and spatially independent profiles of the background
quantities (velocity, density, magnetic field) in the cosmic ray 
precursor (both these assumptions are known to be not
fulfilled, but at present it is not clear what the consequences
of relaxing these assumptions might be).

The results illustrated above are based on two related but separate 
arguments: 1) the particle acceleration at non-relativistic shocks
occurs with a large efficiency; 2) this efficient acceleration is able 
to amplify the magnetic field.

It is easily understandable that this implies a quite non-linear
scenario in which the spectrum of cosmic rays determines the spectrum
of magnetic fluctuations, which in turn determines the diffusive
properties felt by cosmic rays and therefore their spectrum. The 
amplification of the magnetic field also implies that the maximum
energy of the accelerated particles increases, which in turn enhances
the shock modification. 

Despite all these promising developments on the theoretical side,
the direct evidence of acceleration of hadrons at the shocks developed
in SNRs is still missing. The smoking gun could come from the
detection of gamma rays generated in the decay of neutral pions coming
from inelastic nucleus-nucleus collisions. The recent detection of
gamma rays by HESS \cite{hess} might be the long searched signal, 
but this needs further confirmation. 
However, indirect circumstantial evidences of efficient acceleration of
hadrons and of magnetic field enhancement are not lacking. The concave
spectra that are the peculiar feature of acceleration at strongly
modified shocks might be required to explain the multifrequency 
observations of some SNRs \cite{concave}. Although the radiations
are produced by electrons, the latter are actually accelerated in the 
velocity background generated by the dynamical backreaction of
protons, since electrons hardly contribute any pressure when compared
with the hadronic component.

The recent paper by \cite{volk} provides us with a review of the
magnetic fields inferred in several SNRs as obtained from the X-ray
brightness of the shells. X-rays are generated by synchrotron emission
of relativistic electrons, and the extension of the X-ray bright
region depends on the strength of the magnetic field. The stronger the
field the narrower the emitting region. In all cases considered by
\cite{volk} the inferred fields are in the range of a few $100\mu G$,
suggesting a substantial magnetic field amplification, 
possibly induced by cosmic rays. The authors of Ref. \cite{pohl} have
however argued that on the same spatial scale of the variations in the
X-ray brightness a substantial damping of the amplified magnetic field
is also expected. In this case the observed variations in the X-ray
brightness would reflect the damping of the magnetic field rather than 
the energy losses of energetic particles. 

\subsection{Relativistic shocks}
\label{sec:shocks2}

For non-relativistic shocks the high velocity of the accelerated
particles ($\sim$ the speed of light) compared with the shock speed 
makes the particle distribution
quasi-isotropic. This is no longer true for relativistic shocks, where
the shock and the particles share roughly the same speed and there is
not enough time for isotropizing the particles. This fact makes the
acceleration qualitatively and quantitatively different in the case of
relativistic shocks. An initially isotropic distribution of particles
in the upstream fluid is transformed by the shock into a highly
anisotropic distribution due to relativistic beaming, provided the
particles can make their way back from the downstream region. 

\begin{figure}[thb]
 \begin{center}
  \mbox{\epsfig{file=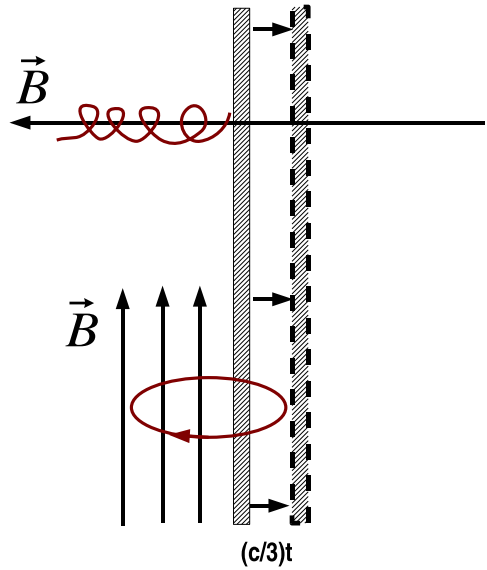,width=11.cm}}
  \caption{Schematic view of particle acceleration at a relativistic 
shock without magnetic scattering agents. The case depicted at the top
(bottom) refers to a magnetic field parallel (perpendicular) to the 
shock normal.}
 \end{center}
\label{fig:relsh}
\end{figure}

In Fig. 4 we plot a schematic view of acceleration at
a relativistic shock front in the absence of scattering agents in the
downstream plasma. The top (bottom) part shows the case of a magnetic field 
parallel (perpendicular) to the shock normal (case I and case II
respectively). In the downstream frame of an ultra-relativistic shock
the shock itself moves with a velocity $\sim c/3$. During the time $t=2\pi
r_L/c$ necessary for relativistic particles to travel one Larmor
rotation, the shock has moved by 
\begin{equation}
\Delta x \sim \frac{1}{3} c \frac{2\pi r_c}{c} = \frac{2\pi}{3} r_L
> r_L.
\end{equation}
This implies that in the absence of magnetic scattering, it is
virtually impossible for particles to re-cross the shock and take part
in the acceleration process. Many different approaches have shown
however that scattering helps in taking the particles back upstream
(see for instance the simulations in Refs. 
\cite{ostro,lemoine,achter,kirkscnum} 
and the analytical approaches by 
\cite{kirkscana,achter,achter1,vietri,bvietri}). 

Contrary to what happens in the case of non-relativistic shocks, the
spectrum of particles accelerated at relativistic shocks is not
universal, though being a power law. The slope of the spectrum 
depends on the details of the scattering of the particles in the 
upstream and downstream regions. 
In particular, if the {\it small pitch angle scattering}
(SPAS) assumption is made, then some kind of universality is recovered 
for large Lorentz factors $\gamma_{sh}\gg 1$, and the slope of the
distribution function in momentum space approaches
$s=4.32$ \cite{achter1,lemoine,bvietri}. 
On the other hand at some sufficiently large value
of $\gamma_{sh}$ the assumption of SPAS is expected to break. At this
point it is expected that the spectrum hardens \cite{bvietri} with
respect to the {\it universal} spectrum. 

The slope is also demonstrated to depend on the equation of state of
the gas downstream, which determines the compression factor at the shock. 
For instance, for a shock speed $u=0.9$ in units of the speed of
light, the spectrum in momentum space has a slope $s=4.71$ for the case 
of the so-called relativistic equation of state $u u_d=1/3$ ($u_d$ here is
the speed of the downstream fluid in the shock frame), but it is 
$s \sim 4.1$ for the probably more realistic Synge equation of state
\cite{synge}. In the left panel of Fig. 5 
(from \cite{bvietri}) we plot the predicted spectral slope as a 
function of the shock compression factor for $u=0.8$ and $u=0.9$. 

\begin{figure}[thb]
 \begin{center}
  \mbox{\epsfig{file=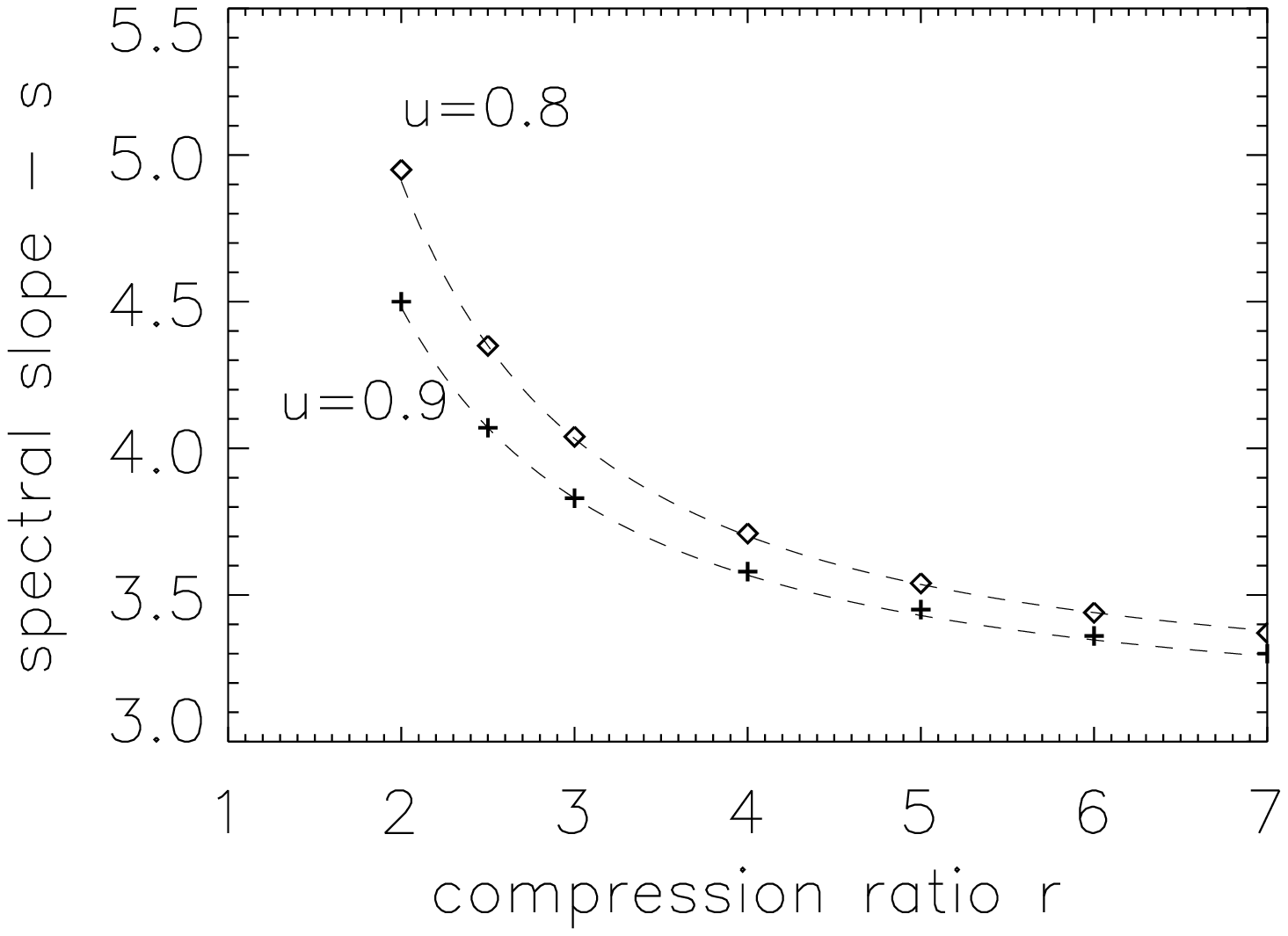,width=6.1cm}}
  \mbox{\epsfig{file=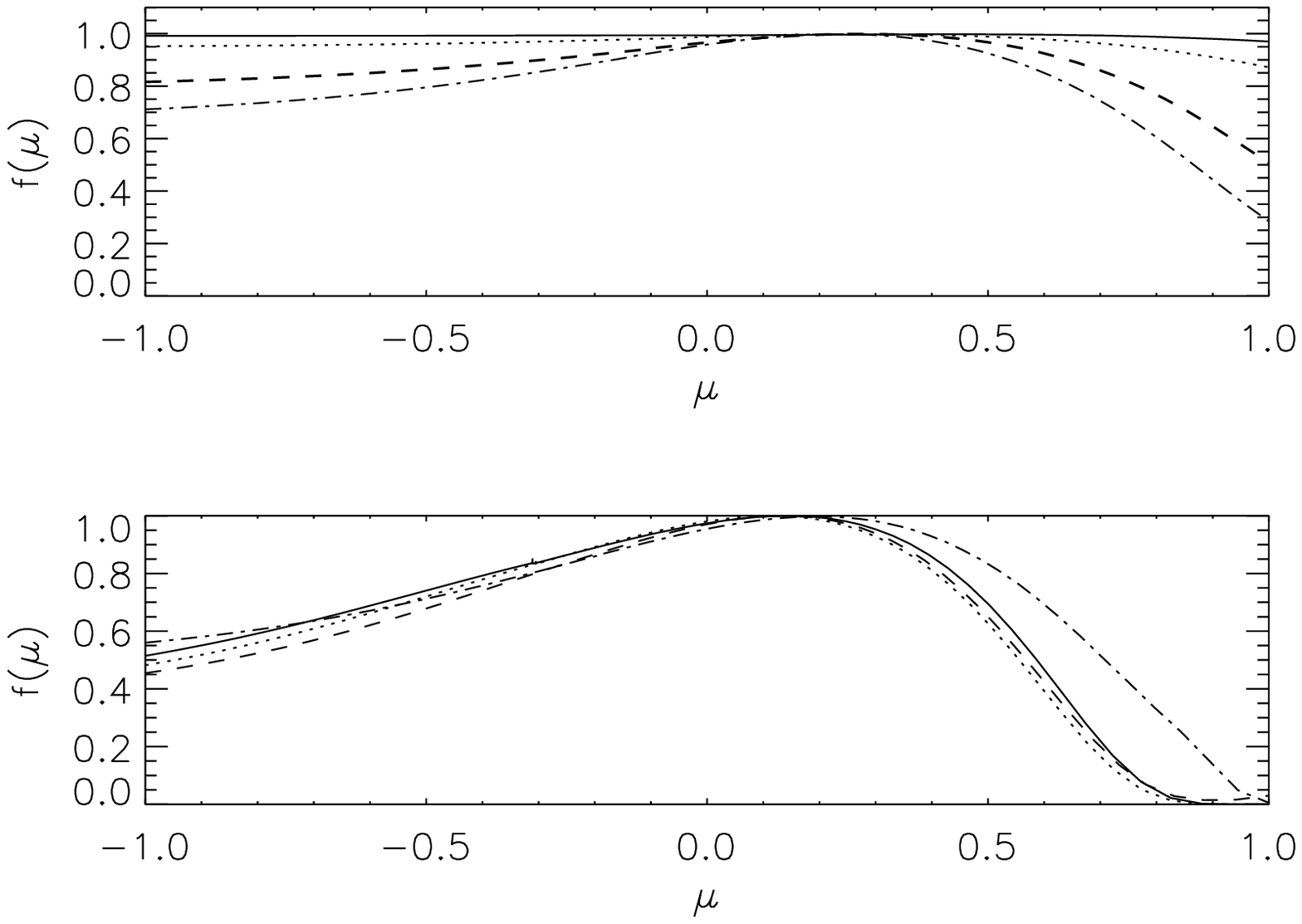,width=6.1cm}}
  \caption{[On the Left]-
Slope of the spectrum of accelerated particles for a shock
with speed $u=0.8$ (diamonds - upper curve) and $u=0.9$ (crosses -
lower curve) in the regime of large angle scattering, as a function
of the compression ratio. The continuous lines are the fit to the 
results of Ref. $^{61}$.
[On the right]-
Upper panel: Distribution function for $\gamma_{sh}\beta_{sh}=0.04$ (solid line),
$\gamma_{sh}\beta_{sh}=0.2$ (dotted line), $\gamma_{sh}\beta_{sh}=0.4$
(dashed line) and $\gamma_{sh}\beta_{sh}=0.6$ (dash-dotted line).
Lower panel: Distribution function for $\gamma_{sh}\beta_{sh}=1$ (dash-dotted line),
$\gamma_{sh}\beta_{sh}=2$ (dashed line), $\gamma_{sh}\beta_{sh}=4$
(dotted line) and $\gamma_{sh}\beta_{sh}=5$ (solid line).}
 \end{center}
\label{fig:slope_and_ani}
\end{figure}

As qualitatively explained above, the distribution function of
particles accelerated at shocks moving at speed close to the speed 
of light is expected to be anisotropic, both upstream and downstream.
In the right panel of Fig. 5 (from
\cite{bvietri}) we show the downstream distribution function for
the shock speeds listed in the caption, as calculated in Ref. 
\cite{bvietri}. It is clear how the anisotropy increases when the
shock becomes increasingly more relativistic. The slopes in momentum
space for the cases considered in Fig. 5 (right panel) 
are listed in Table 1 \cite{bvietri}.
\begin{center}
\begin{tabular}{|c|c|c|c|} \hline
$\gamma_{sh}\beta_{sh}$ & $u$ & $u_d$ & $slope~s$ \\ \hline
  0.04 &   0.04  & 0.01  & 4.00 \\
    0.2 &   0.196 & 0.049 & 3.99 \\
    0.4     &   0.371 & 0.094 & 3.99 \\
    0.6 &   0.51  & 0.132 & 3.98 \\
    1.0 &   0.707 & 1.191 & 4.00 \\
    2.0 &   0.894 & 0.263 & 4.07 \\
    4.0 &   0.97  & 0.305 & 4.12 \\
    5.0 &   0.98  & 0.311 & 4.13 \\
\hline
\end{tabular}
\end{center}

The phenomenon of shock modification induced by cosmic rays has 
received much less attention for the case of relativistic shocks
(see \cite{baring} and \cite{double}).

\section{The transition from galactic to extragalactic origin: an
  ankle or a dip?} \label{sec:transition}

The structure formed by the second knee and the dip, illustrated in 
Fig. 1, has been traditionally interpreted as the 
transition from galactic to extragalactic cosmic rays, and named 
{\it the ankle}. This rather sharp feature would be the result of the
superposition of a rapidly falling galactic spectrum and a rising 
(in the $E^3 J(E)$ formalism) spectrum of extragalactic cosmic rays.
This scenario requires that the cutoff in the galactic component 
exceeds $10^{19}$ eV. 

More recently an alternative interpretation has been put forward, with very
significant implications for the origin of cosmic rays at large. 
In \cite{bere1,bere2} it was pointed out that the combination of pair
production energy losses and adiabatic energy losses, due to the
expansion of the universe, would generate a spectrum of cosmic rays at
the Earth with a feature which fits the observed second knee
and dip (see Fig. 6 for predictions obtained following 
the calculations of \cite{beregrigo,bere1,bere2}). We will refer to this model
as the {\it twisted ankle} (TWA) scenario. In this model the transition
from galactic to extragalactic cosmic rays would take place at
energies below $\sim 10^{18}$ eV, where the galactic cosmic rays
would be cut off. Much larger maximum energies of galactic cosmic
rays are required in the {\it ankle} scenario, where the galactic
component disappears only at $E\geq 10^{19}$ eV. In its basic form, 
the TWA scenario implies that cosmic rays are injected at extragalactic
sources with an injection spectrum $E^{-2.7}$ and no luminosity
evolution (solid line in Fig. 6). Although it is a 
worse fit, the data appear to show a dip-like structure also for the
case of evolving sources, with injection spectrum $E^{-2.4}$ and
luminosity evolution that scales with the redshift $z$ of the sources
as $L(z)\propto (1+z)^4$ (dashed line in
Fig. 6). In both cases, even small levels of
magnetization of the intergalactic medium would induce a low energy
suppression of the flux, which might in fact improve the fit to
the all-particle spectrum \cite{lem,alo}.  

\begin{figure}[thb]
 \begin{center}
  \mbox{\epsfig{file=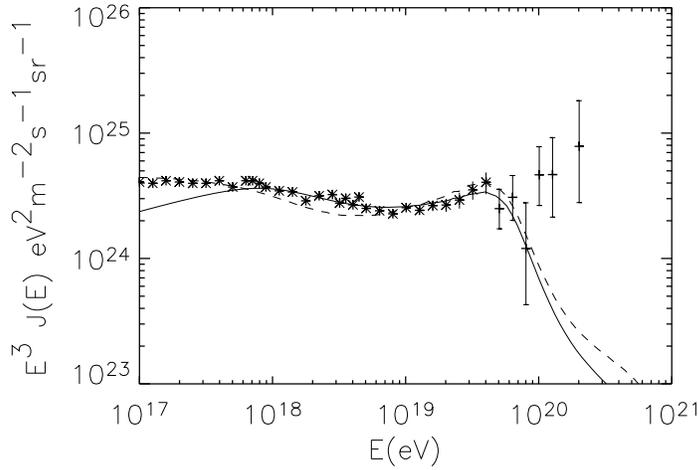,width=10.cm}}
  \caption{Spectrum of cosmic rays of extragalactic origin for an
  injection spectrum $\propto E^{-2.7}$ and no luminosity evolution 
  (solid line) and for injection spectrum $\propto E^{-2.4}$ and
  luminosity evolution $L(z)\propto (1+z)^4$ (dashed line). The data 
  are from Akeno and AGASA.}
 \end{center}
\label{fig:27vs24}
\end{figure}

The most appealing aspects of the TWA scenario can be summarized as
follows: 1) the dip is a natural consequence of particle physics: it
is simply produced by the combination of abiabatic energy losses and
Bethe-Heitler losses. 2) the model does not require galactic sources to
accelerate cosmic rays to energies in excess of $10^{19}$ eV, which
would represent a problem for the vast majority of acceleration
models applied to galactic sources (see however \cite{epstein,arons} for
a model in which acceleration of nuclei to such energies is achieved,
although with too flat a spectrum to be relevant for this physical 
situation).

If the rigidity model suggested by the KASCADE data and discussed
in Sec. \ref{sec:shocks} is correct, it is not easy to argue in 
favor of a heavy nuclear component extending to energies $>10^{19}$ 
eV. Although this point is rather weak at the present time, because
of the large uncertainties in the observations, it might become a
more solid point as more reliable measurements of the chemical
composition and cosmic ray spectrum in this energy region become
available. It is crucial that future experiments give priority to
the measurement of the chemical composition in the energy region 
between $10^{17}$ eV and $10^{19}$ eV.

The relatively steep injection spectrum required by the TWA scenario
may be problematic in that it appears to contradict the common wisdom that
strong shock waves accelerate particles to flat spectra (typically
$E^{-2}$). However, as discussed in Sec. \ref{sec:shocks}, this is a
rather oversimplified conclusion as both non-relativistic shocks and
relativistic shocks can accelerate particles with steeper (as well 
as flatter) spectra in
realistic situations: for instance a shock wave with Mach number 2.6 
generates a power law with slope 2.7, although the acceleration
efficiency is probably rather low, due to the low value of the Mach
number. For relativistic shocks, the spectrum of accelerated
particles, as discussed in Sec. \ref{sec:shocks}, may depend on the 
equation of state. In Sec. \ref{sec:shocks2} we showed an instance 
in which a shock with speed
$u=0.9$ generates accelerated particles with slope 2.7. Clearly none
of these situations is universal, but these cases serve the scope of 
pointing out that this alleged discrepancy between the slope required 
by the TWA model and the expectation from the theory of shock acceleration
cannot be taken too seriously.
Moreover, both luminosity evolution of the sources (dashed line in
Fig. 6), and the overlap of the contribution from 
numerous sources with different maximum energies at the source 
\cite{michael} can produce a dip in the diffuse cosmic ray spectrum 
and at the same time may require relatively flat injection spectra.

A more serious concern for the TWA model is related to the chemical 
composition: the dip is in fact likely to disappear if a small
contamination (with solar-like abundance) of nuclei heavier than
hydrogen is present at the source \cite{bere2,angela}. In this respect
Helium appears to be the most {\it dangerous} in terms of affecting
the conclusions of the model. Magnetic horizon effects related with 
nuclei with different charge to mass ratio might somewhat
mitigate the relevance of this issue \cite{sigl}. 

\section{Spectrum and small scale anisotropies of UHECRs}\label{sec:gzk}

A solid prediction of particle physics is that the photopion reactions 
of protons on the cosmic microwave background during their journey
from the sources to Earth in the intergalactic space should induce a
suppression in the diffuse spectrum of cosmic rays. This suppression, 
usually referred to as the GZK feature \cite{gzk}, appears at energy
$\sim 10^{20}$ eV as a consequence of the relatively short interaction 
length, large inelasticity of the reaction of photopion production 
and due to the fact that the reaction has a kinematic threshold at
roughly this energy (in fact the reaction starts taking place for 
lower energy protons when they scatter against the CMB photons 
on the tail of the Planck distribution, but most interactions occur with 
the photons on the peak). In Fig. 7 we show the loss length as a 
function of the energy of protons. We see that: {\it a)} 
the loss length at $10^{20}$ eV (slightly above threshold for 
$p+\gamma \to \pi + \rm anything$) is $\sim 100$ Mpc while at 
energies two times smaller (slightly below threshold) it is almost 
as large as the size of the universe
(horizontal line). The corresponding flux of cosmic rays at
these energies is expected to drop roughly by the ratio of the 
loss lengths if the sources have a spatially homogeneous distribution.
The exact amount of the suppression is however quite
sensitive to the injection spectrum, to the maximum energy at the
source, to the redshift evolution of the sources and to the spatial 
distribution of the sources; {\it b)} 
at energy $\sim 2\times 10^{18}$ eV the loss length, dominated by the
process of proton pair production, equals the loss length due to the 
expansion of the universe (abiabatic losses). 

\begin{figure}[thb]
 \begin{center}
  \mbox{\epsfig{file=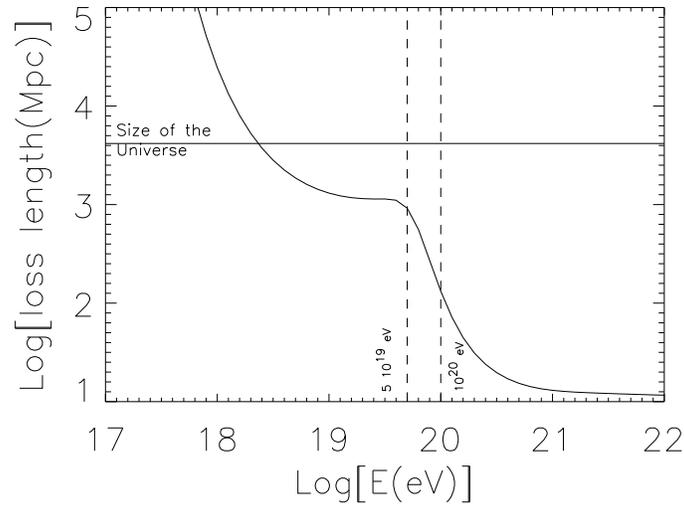,width=11.cm}}
  \caption{Combined loss length for Bethe-Heitler pair production and
  photopion production on the photons of the cosmic microwave
  background. The horizontal dashed line shows the size of the
  universe and therefore the onset of adiabatic energy losses.}
 \end{center}
\end{figure}

While the first point (point {\it a}) gives rise to the GZK feature, the second
(point {\it b}) determines the appearance of the dip, discussed in 
Sec. \ref{sec:transition} \cite{bere1,bere2}.

The very low number of events expected at energies around and
above $10^{20}$ eV makes the detection of the GZK feature very problematic. 
Experiments that have operated so far have not been successful 
in either detecting the GZK feature or proving its absence with a
sufficiently high statistical significance. The
spectra of AGASA, HiRes and the newly released data from the
Pierre Auger telescope \cite{augerdata} are plotted in Fig. 8.

In \cite{dbo1} a careful
Monte Carlo simulation of the propagation of cosmic rays from
homogeneously distributed sources allowed the authors to infer the
statistical significance of the AGASA and HiRes data available at the
time. Neither one of the two experiments was found to have reached a
definitive conclusion on the presence or absence of the GZK feature.
AGASA and HiRes have a systematic offset in the absolute flux 
normalization that can however be understood if a relative systematic
error in the energy determination of $\sim 30\%$ is assumed.

\begin{figure}[thb]
 \begin{center}
  \mbox{\epsfig{file=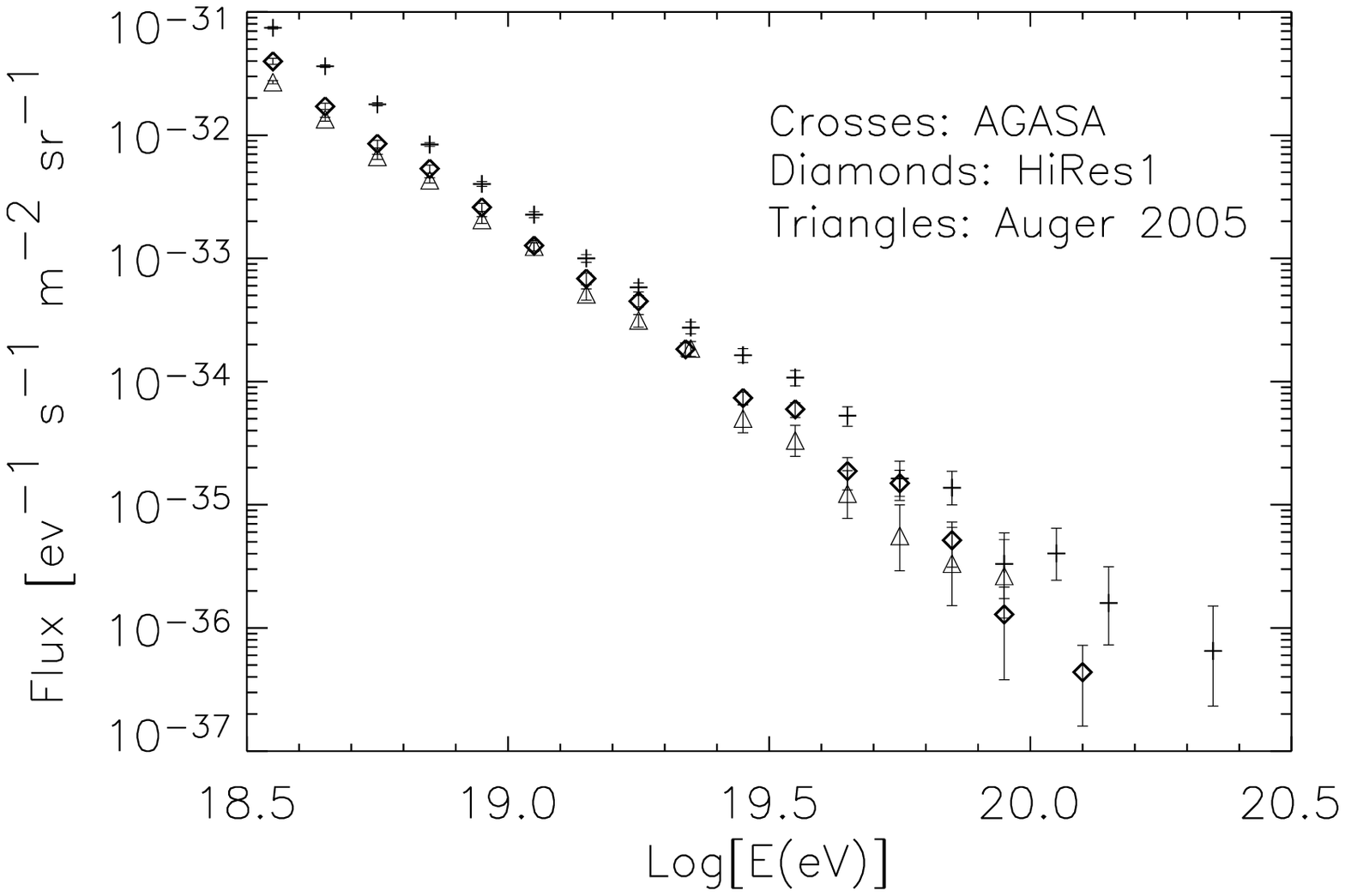,width=6cm}}
  \mbox{\epsfig{file=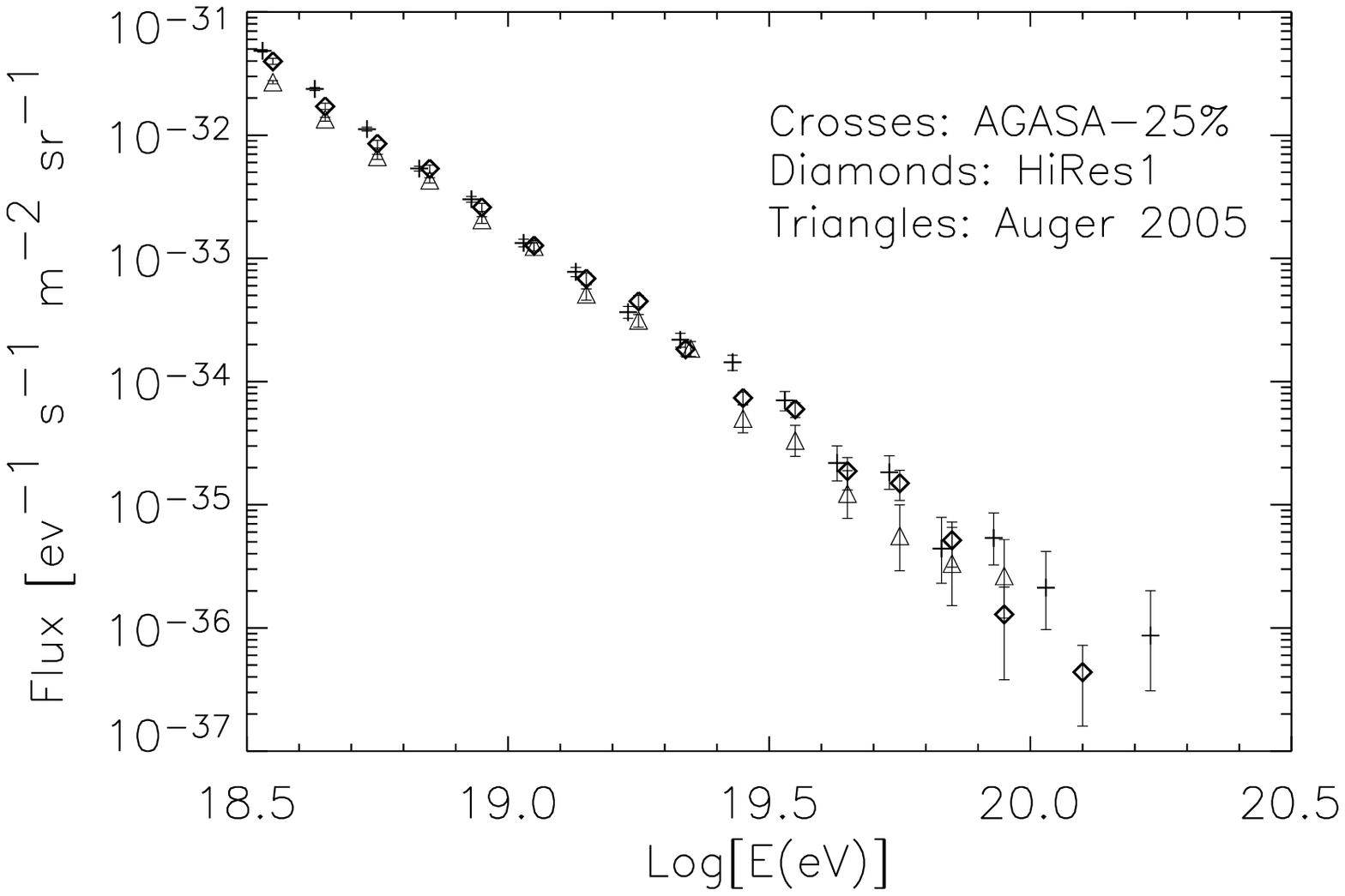,width=6cm}}

  \caption{{\it Left Panel)} Spectrum of AGASA, HiRes and Auger. 
{\it Right Panel)} Spectrum of AGASA, HiRes and Auger with a
correction of the AGASA energy for a possible systematic error in the 
energy determination by $-25\%$.}
 \end{center}
\label{fig:data}
\end{figure}

The results of \cite{dbo1} were recently confirmed and discussed in
more detail in \cite{dbo2}. The authors simulated the AGASA spectrum 
normalizing at $4\times10^{19}\rm eV$ with a 30\% statistical error in
the energy determination of the events. The simulations were
ran for 30000 realizations of spectra with 72 events above
$4\times10^{19}\rm eV$ like in the AGASA data and the number of events 
with energies above $10^{20}\rm eV$ in each realization were
recorded. Taking into account the crucial non-gaussian distribution
of the simulated data, the simulations show that the probability of
having 11 or more events (as in AGASA) is $6\times10^{-4}$. In terms
of gaussian probabilities this would correspond to about $3.2\sigma$.
The role of systematic errors was also investigated in \cite{dbo2}:
the authors conclude that the AGASA data are compatible with the
presence of a GZK feature at the $2.5\sigma$ level (in terms of
gaussian errors). A similar exercise carried out on the HiRes data
also results in the conclusion that HiRes data are {\it away} from
AGASA data only at the $2\sigma$ level \cite{dbo2}.

These statistical considerations are the result of averages over large
samples of realizations of source distributions, therefore one might
wonder whether the individual spectra of those realizations which have a
large number of events at ultra high energies are similar to the AGASA 
spectrum or not. In Fig. 9 we plot the spectra of some of
the realizations that showed 11 or more events above $10^{20}\rm eV$ 
(the error bars here are just poissonian, namely they have
the same meaning as in Fig. 8). These 
spectra closely resemble the AGASA spectrum and all of them show no
evidence of a GZK suppression, despite the fact that in the lower energy
region they all fit the data quite well. This shows that an AGASA-like
spectrum is not that improbable, even if the {\it average} cosmic ray
spectrum can be expected to show a GZK feature \cite{dbo2}.

\begin{figure}[thb]
 \begin{center}
  \mbox{\epsfig{file=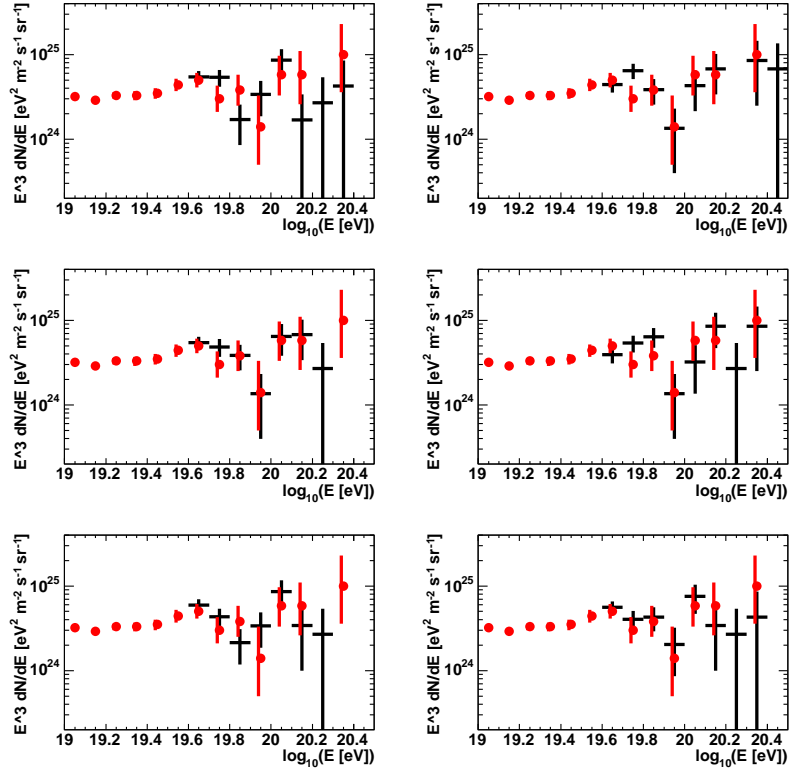,width=11.cm}}
  \caption{Some of the simulated realizations with number of events
  equal to or larger than the actual number detected by AGASA $^{78}$. }
 \end{center}
\label{fig:realiz}
\end{figure}

The spectrum of cosmic rays by itself does not contain enough 
information to determine the type of sources of UHECRs.
A first hint at the type of sources can come from the
identification of small clusters of events with arrival directions
within the error box of the experiment. Such a signal of small scale
anisotropies (SSA) was first claimed by the AGASA collaboration 
\cite{agasa_ssa}. The signal was however shown to have a 
low statistical significance \cite{finley}. On the other 
hand, if astrophysical point sources are believed to accelerate
UHECRs and if the intergalactic magnetic field is sufficiently low,
then SSA are to be expected. It was shown in Ref. \cite{danny} that, if
the AGASA results were confirmed, the number density of sources could be
estimated to be around $10^{-5}\rm Mpc^{-3}$, with a quite large
uncertainty due to the very limited statistics of events available
(see \cite{allssa} for other estimates). In \cite{dbo2} the authors
show that the spectrum of AGASA appears to be not fully consistent
with the detection of the SSA by the same experiment. The possibility
of coupling the information on the spectrum and SSA to gather better 
information on the number of sources will probably prove useful with
the upcoming data from the Pierre Auger Collaboration. 

The role of intergalactic magnetic fields in changing the spectrum and 
clustering of UHECRs is badly constrained as different simulations
give different estimates for the magnitude and spatial structure of 
these fields (see, e.g., \cite{dolag04,dolag05,siglsim1,siglsim2}).
A homogeneous magnetic field spread through the whole Universe would leave
the spectrum of diffuse cosmic rays unchanged with respect to the case
of the absence of magnetic field \cite{aloprop}. 
The intergalactic magnetic fields can be neglected for particles with
energies above $4\times 10^{19}$ eV if the magnetic
field in the intergalactic medium is less than $\sim 0.1~\rm nG$ 
(reversal scale  of $1$ Mpc) and the small scale anisotropies are
evaluated on angles of $\sim 2$ degrees \cite{danny}. This
field magnitude is compatible with observational bounds \cite{bbo99} 
and detailed numerical simulations \cite{dolag04,dolag05} 
(however, see \cite{siglsim1,siglsim2} for
different results). In these numerical simulations the magnetic 
field is assumed to be formed at some early time (seed field) and later
amplified during the formation of the large scale structure of the
Universe. The initial seed field can be due to some battery mechanism
at shock fronts formed at the turn-around surfaces of large scale
structures or may be of primordial origin \cite{dariorev}. It
might also have been spread into the universe during
the early stages of evolution of galaxies \cite{lesch}. In many of
these cases however the resulting magnetic fields are rather model
dependent and are affected by further evolution of magnetic fields in
the surrounding medium. In particular, the evolution due to the
formation of large scale structures as found in simulations affects
all these scenarios. An upper limit on the initial seed field could be
imposed by requiring that the cosmological evolution of the fields
does not lead to exceed the strength of the magnetic field seen in 
some clusters of galaxies (a few $\mu G$).

As pointed out above, the statistical significance of the SSA found
by AGASA is rather low, and quite larger statistics of events are 
needed in order to detect the small angle clustering, if present. 
These measurements need to be carried out at extremely high energies 
in order to avoid appreciable deflections by the galactic magnetic
fields. No evidence of SSA has been found so far in the HiRes data
\cite{hires_ssa}. 

The search for the GZK feature in the spectrum of UHECRs has been 
for a long time the prominent goal of researchers in this field. 
One should however not forget that there is an independent challenge,
which is that of identifying sources that can potentially host an
accelerator able to energize particles to energies as high as 
$\sim 10^{20}$ eV or larger \cite{angelarep}. From the theoretical point of
view, the most important development in the field of particle 
acceleration has been the possibility of strong magnetic field 
amplification at shocks, as discussed in Sec. \ref{sec:shocks}. This
mechanism however has currently received much attention only in 
connection to SNRs, where the highest energies that can be reached 
are much lower that $\sim 10^{20}$ eV. 

From the observational point of view, the most problematic issue to
explain is the lack of any counterparts to the Fly's eye event 
\cite{fly}, with energy $3\times 10^{20}$ eV. The distance of the
potential sources should be limited to roughly the loss length 
of protons with this energy (even less in the case of gamma rays),
namely, from Fig. 7, $\sim 20$ Mpc. On such short 
distances the effect of magnetic fields in deflecting the particles
from their direction, if any, should be negligible, and a clear 
identification of the source should therefore be possible. No
candidate was instead found \cite{sommers}. This problem is still
unsolved, though several possibilities have been discussed. Among
these the most likely is probably that the source was {\it bursting}.
The term {\it burst} here is used to indicate a phenomenon with 
duration shorter than the typical time delays induced by magnetic
fields either in the intergalactic medium or in the Galaxy itself. 

\section{Summary and Discussion}
\label{sec:summary}

Understanding the origin of cosmic rays means providing an explanation
for the acceleration and the propagation of these particles, for their
chemical composition and for the way they generate secondary
radiations that we end up observing either as sources or as diffuse
radiation in the Galaxy (or may be in the intergalactic medium). 

In the last few years we have moved significantly ahead in terms of
achieving these goals: the KASCADE data strongly suggest that the
all-particle spectrum is the result of different chemical components
with rigidity dependent maximum energies. Each chemical component
appears to have a pronounced knee where the slope of the observed
spectrum changes rather drastically (opposite to the change of slope
in the all-particle spectrum, which is instead rather small). It remains
to be understood whether these knees are the result of propagation
in the Galaxy or of the acceleration process. The KASCADE rigidity
dependent knees suggest that the galactic cosmic rays should {\it end}
at energies $\sim 10^{18}$ eV with a presumably Iron dominated 
chemical composition. This energy is quite smaller than the position 
of the ankle, so that the confirmation of this result would imply that
the transition from a galactic to an extragalactic origin of cosmic
rays should take place below the ankle. This would rule in favor of
the alternative view that the region of the ankle, made of a second
knee and a dip is the result of Bethe-Heitler pair production
on extragalactic cosmic rays \cite{bere1,bere2}, although at the
present time we are far from having proved the model right. In order 
to clarify this problem it is crucial that the chemical composition 
of cosmic rays between $10^{17}$ eV and $10^{19}$ eV is reliably
measured. If the extragalactic cosmic ray spectrum is appreciably 
contaminated by heavier elements the TWA model of the transition also 
has problems \cite{angela,bere2}.

From the theoretical point of view, after the proposal of
\cite{bl00,bl01,bell2004}, it seems that the acceleration of cosmic 
rays at the shock fronts of SNRs, together with the high acceleration
efficiency of such shocks, may make the acceleration of protons (iron 
nuclei) up to a few $10^{16}$ eV ($10^{18}$ eV) viable. It remains to
be understood why we do not observe enough SNRs in the TeV range
(see \cite{hillas} and references therein for an accurate discussion 
of this point). 

Moving toward higher energies, the spectrum of cosmic rays with energy
between $3\times 10^{18}$ eV and $5\times 10^{19}$ eV as measured by 
AGASA (and Akeno), HiRes and now Auger (listing only the experiments 
with comparable exposures at the present time), seems to be rather
well determined. There is an offset in the absolute normalizations of 
the three experiments which however could be resolved by admitting a 
systematic error in the error determination (roughly $30\%$) by one 
or all of the three experiments, possibly reflecting different 
observational techniques. Unfortunately this small difference implies
different conclusions as far as the detection of the GZK feature is
concerned. This is the clear symptom of an insufficient statistics of
events, as shown quantitatively in \cite{dbo1,dbo2}. 

A signal associated with small angle clustering of the arrival
directions, initially claimed by the AGASA collaboration
\cite{agasa_ssa} also appears to be statistically shaky
\cite{finley}. Both these points should strongly drive the 
community to achieve the larger statistics of events necessary
for a reliable measurement of the spectrum and for a careful 
identification of anisotropies, two preliminary steps to have
a clue to the sources of ultra high energy cosmic rays. This
drive is leading us to the Auger Telescope and will hopefully
lead us to the detection of UHECRs from space, with EUSO-like
\cite{EUSO} or OWL-like experiments \cite{OWL}.

\section*{Acknowledgments}

The author is very grateful to R. Aloisio, E. Amato, V. Berezinsky, 
D. De Marco, D. Ellison and S. Gabici for constructive comments on 
the manuscript and ongoing collaboration. The authors also wishes 
to acknowledge the help of T. Stanev and J. H\"{o}randel for providing 
different versions of Fig. 1. Finally, the author is grateful for
the kind hospitality of KIPAC at SLAC and at Stanford University
in october-november 2005.

\end{document}